# Effect of inelastic ion collisions on low-gain avalanche detectors explained by an $A_{Si}$-$Si_i$-defect mode

Kevin Lauer[1,2], Stephanie Reiß[1], Aaron Flötotto[2], Katharina Peh[2], Dominik Bratek[2], Robin Müller[2], Dirk Schulze[2], Wichard Beenken[2], Erik Hiller[1], Thomas Ortlepp[1], Stefan Krischok[2]

[1] CiS Forschungsinstitut für Mikrosensorik GmbH, Erfurt, Germany
[2] TU Ilmenau, Institut für Physik and Institut für Mikro- und Nanotechnologien, Ilmenau, Germany

Abstract

The acceptor removal phenomenon (ARP), which hampers the functionality of low-gain avalanche detectors (LGAD), is discussed in frame of the $A_{Si}$-$Si_i$-defect model. The assumption of fast diffusion of interstitial silicon is shown to be superfluous for the explanation of the $B_{Si}$-$Si_i$-defect formation under irradiation, particular at very low temperatures. The experimentally observed properties of the ARP are explained by the donor properties of the $B_{Si}$-$Si_i$-defect in its ground state. Additionally, low temperature photoluminescence spectra are reported for quenched boron doped silicon showing so far unidentified PL lines, which change due to well-known light-induced degradation (LID) treatments.

## 1. Introduction

When many particles with high energy impinge on well-ordered materials like crystals the regime will be damaged up to a complete loss of order, which is called amorphization. A new order or crystal structure with less defects can be rebuilt by subsequent careful and sophisticated annealing. A long history of approaches to understand the damage to optimize device functionality does exist. In particular the semiconductor material silicon is heavily investigated.[1] Nevertheless, the current profound knowledge of silicon has not led to fully understanding and circumventing several recent problems like the light-induced degradation (LID)[2] in silicon solar cells or the acceptor removal phenomenon (ARP) in low-gain avalanche detectors (LGAD).[3]

Detectors, which feature an internal gain such as avalanche photodiodes (APD), are well known for several decades.[4] In recent years it was suggested to tailor the gain of these devices to a low value to enhance their noise and timing properties.[5] These devices are called LGADs and are manufactured by several institutes like the CiS Forschungsinstitut für Mikrosensorik GmbH (CiS). These well working devices have one drawback that limits their application in high luminosity ambience. The gain disappears after some time under operation in a heavy irradiation environment. This was attributed to a removal or deactivation of the boron acceptor within the gain layer of a LGAD.[6]

This behavior of gain removal after irradiation was also verified by us in a recent investigation where the disappearance of the gain in the LGAD developed at CiS was observed.[7] The gain reduction of the LGAD after proton irradiation found in this recent investigation is reprinted in Figure 1.

One tentatively identified defect responsible for the ARP is the $B_iO_i$ defect.[3] In frame of that model a long range migration of boron at room temperature is necessary for this defect formation process under irradiation. The main outcome of the preceding collaborative work of CiS and TU Ilmenau was that a possible long-range migration of boron atoms under irradiation could not be visualized by SIMS measurements in a specifically designed experiment. This finding suggests the suspicion that the boron atoms do not move through the lattice but rather boron captures a silicon interstitial forming the $B_{Si}$-$Si_i$-defect.

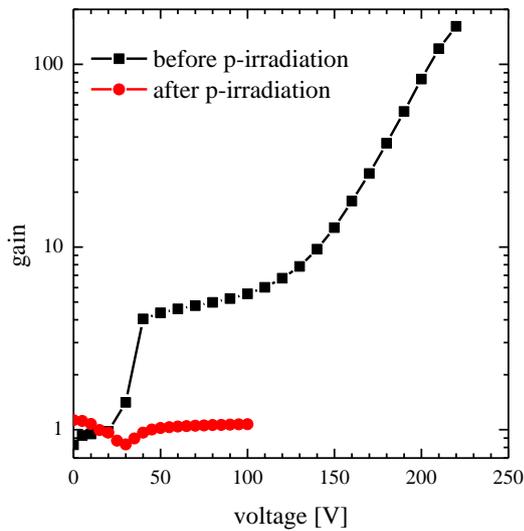

**Figure 1: Gain of a LGAD developed at CiS as a function of the reverse bias voltage before and after proton irradiation. Reprinted from Ref. [7]**

In frame of the acceptor removal phenomenon observed in LGADs an interesting experimental work of Buzynin et al. should be discussed here.[8] In this study the propagation of type inversion of moderately boron doped silicon was investigated while low-energy Ar irradiation of one side of a silicon wafer. A deactivation of boron acceptors was observed, which was accompanied by the creation of a n-type region. The interpretation of the experimental results within the study is as follows: Argon ion irradiation leads to silicon interstitial generation near the silicon wafer surface due to the low incidence depth of the argon ions. These silicon interstitials migrate at about 100°C through the wafer.[8] A diffusivity of the silicon interstitials at this temperature of about $10^7 cm^2/s$ could be estimated from this experiment. During their journey through the wafer the silicon interstitials are captured by boron acceptors. The resulting defect loses its acceptor properties and eventually becomes even a donor. But the donor character of the resulting defect could not be proved by that investigation. Finally, after most of the boron atoms have captured a silicon interstitial atom, the silicon becomes n-type. In the work of Buzynin et al.[8] several ideas to explain this acceptor removal including fast migration of oxygen are discussed, but a convincing solution could not be provided.

The understanding of the fate of the silicon interstitial in silicon is very limited. This fact becomes visible in the book by Pichler[9] who recapitulates diffusion coefficients of many authors, which believed to measure this quantity. The span of reported diffusion coefficients of interstitial silicon at e.g. 900°C is nearly 10 orders of magnitude. Similar problems are found for the interstitial silicon solubility. The consequence is that a prediction of the behavior of the silicon interstitial in an experiment from literature data is currently impossible.

In this contribution, we present recent low-temperature photoluminescence results of quenched boron doped silicon and mainly discuss some shortcomings in the understanding of irradiation defects in silicon. We show how a mode of the $A_{Si}$-$Si_i$-defect[10] is able to explain the acceptor removal phenomenon in LGADs.

## 2. Experimental

In our investigations we follow closely the understanding of the behavior of interstitial iron in silicon.[11] Short high temperature annealing steps lead to a homogenous distribution and high concentration of interstitial iron in silicon. A fast quenching step after the annealing guaranties that most of the iron is frozen in the interstitial state. After such procedures defect reaction properties of interstitial iron can be analyzed.[12]

To investigate $A_{Si}$-$Si_i$-defects and further understand the ARP, we generate such defects by high temperature annealing with subsequent fast quenching. Such a procedure was found to enhance the intensity of the P line of low-temperature photoluminescence (LTPL) spectra in indium doped silicon drastically.[13] Most likely the solubility of silicon interstitials is increased by the high temperature anneal and the fast quenching prevents the silicon interstitial from recombination with a vacancy. As a consequence, silicon interstitials are frozen in and can form defect complexes with acceptor atoms, like boron or indium.

LTPL spectroscopy is a powerful tool to investigate defects since a luminescence signal is characteristic of a defect in a specific configuration. For a recent description of the used setup see Ref.[14]. In indium doped silicon a PL feature called P line could be linked to a specific state of the $In_{Si}$-$Si_i$-defect.[15] According to a first report on a PL signal also for the $B_{Si}$-$Si_i$-defect,[16] which is possibly also responsible for light-induced degradation (LID) in silicon,[10] it was found that such a PL signal could not be reproduced.[14]

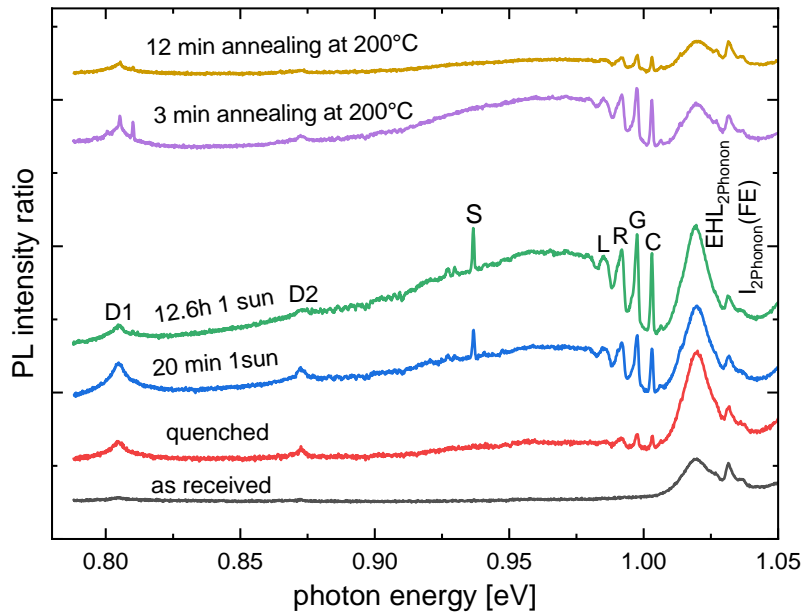

Figure 2: Low-temperature photoluminescence spectra measured on boron doped silicon after quenching and applying LID treatments. PL intensity is normalized to the $I_{2Phonon}(FE)$ peak and shifted for clarity by a constant factor for each spectrum.

Nevertheless, in LTPL spectra of quenched[17] boron doped silicon characteristic PL lines can be found, which exhibit intensity changes due to LID treatments. For this experiment we used boron implanted silicon. The substrate material was FZ silicon with <111> orientation and >10kΩcm boron base doping. The wafer thickness was 400 µm. Before implantation an oxidation step was made revealing 100 nm thick oxide layers on both wafer surfaces. Boron was implanted into the front side of the wafer with an energy of 102 keV and a dose of $4.9 \times 10^{12}$cm$^{-2}$. After implantation, a rapid thermal process (RTP) for 10 s at 1000 °C was applied to activate the boron acceptors. After RTP, the 100 nm thermal oxide was etched off. Subsequently, the wafer was broken into pieces. This status of the samples is denoted by "as received" in Figure 2.

Next, a so-called quenching step was applied. Using tweezers, a wafer piece was held into a flame of a Bunsen burner. After the silicon started to glow red it was held for further 10 s into the flame. Then the piece of silicon was removed from the flame and quenched immediately in ethylene glycol. This status of the sample is denoted by "quenched" in Figure 2.

Then a typical LID treatment[18], which consist of illumination with a halogen lamp for varying times at 1 sun intensity and 40°C, was applied and the LTPL spectra were measured after each treatment step. Finally, a typical LID recovery annealing step at 200°C for several minutes in darkness was applied and the LTPL spectra were measured again.

These spectra are depicted in Figure 2. Every spectrum is normalized with respect to the $I_{2Phonon}$(FE) peak[19] and shifted in the y direction for clarity. In the PL a new line, which is to our knowledge unknown so far, is observed and will be called S line ($E = 0.936$ eV) in the following. This S line appears after quenching and applying a LID treatment by illumination. After annealing at 200°C for 3 min it disappears again. There are some similarities in the behavior of this S line compared to the P line in indium doped silicon: The P line also increases with illumination time. However, in contrast to the P line it disappears for long illumination times.[15] Additionally, a feature called L line ($E = 0.985$ eV) together with its daughter features R line ($E = 0.991$ eV), G line ($E = 0.997$ eV) and C line ($E = 1.003$ eV) are found, which change during the LID treatments. Since, the LID is a complex process[20] further investigation of these lines is clearly necessary.

### 3. Discussion

To discuss the acceptor removal phenomenon in frame of the $A_{Si}$-$Si_i$-defect model some experimental results from the literature will be recapitulated and a very fundamental assumption related to the fate of silicon recoil atoms during irradiation will be reconsidered. Investigation of boron doped silicon irradiated at room temperature is often done using deep level transient spectroscopy (DLTS).[21] These electrical measurements reveal a transient change in capacitance of a n-p-junction after e.g. forward bias pulses which is interpreted to be caused by a removal of negatively charged or neutral defects from the space charge region due to emission of electrons into the conduction band. This change in capacitance for boron doped and irradiated silicon is usually observed around a temperature of 120K.[21] Based on the measured time constants of the capacitance transient at different temperatures an activation energy of the assumed electron emission process can be calculated. The activation energy of this process is often written as a difference e.g. to the conduction band, indicating that the electron has to surmount this barrier to reach the conduction band. In our case we have an activation energy of 0.25eV,[21] which is written as $E_C$-0.25eV. We denote the defect exhibiting this electron emission process as E1. A rigorous discussion of the DLTS method and applicability compared to other methods was recently elucidated by Juhl et al.[22] The advantage of this measurement method is that a measured activation energy is unique to an emission process, which can be correlated with a unique defect. Additionally, the defect concentration can be directly measured by the capacitance change before and after the emission process. Unfortunately, the DLTS peaks have a quite large FWHM making discrimination between closely neighboring activation energies difficult.

Early ideas to assign this observed activation energy after irradiation of silicon to a defect in silicon were made by proposing an interstitial boron-interstitial oxygen ($B_iO_i$) defect due to the observation of the abundance of both species in their investigated silicon.[23] This early assignment was later strongly questioned by investigating intentionally oxygen and boron doped silicon. No correlation of the generation rate of the corresponding defect with the oxygen concentration was found at all.[21] Nevertheless, the assignment of this activation energy measured by DLTS to an assumed $B_iO_i$ defect is still persistent in the literature. Consequently, after investigating the ARP on boron doped test samples by DLTS and finding this activation energy in these samples the defect E1 possibly responsible for the ARP was tentatively identified as $B_iO_i$.[3]

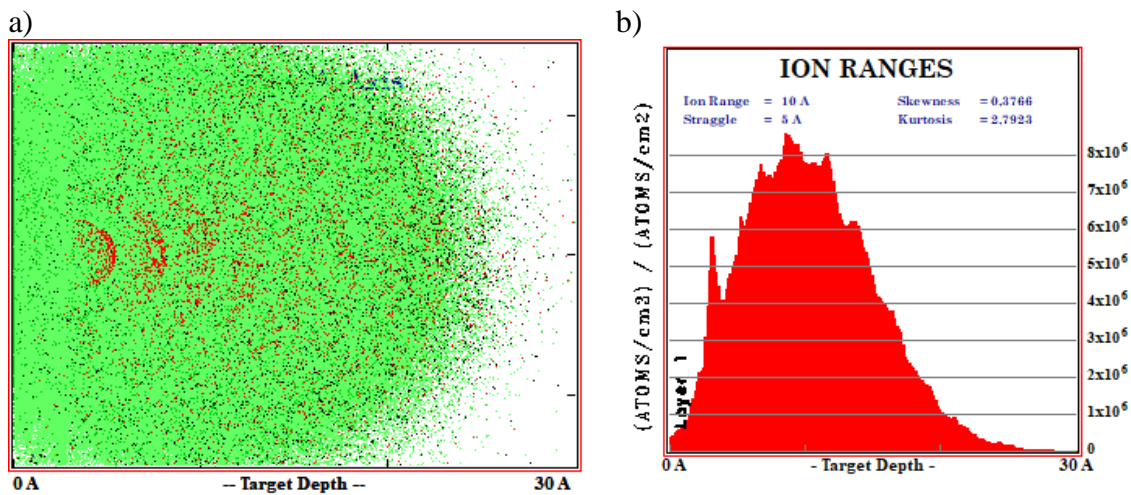

**Figure 3: SRIM simulation of silicon ions with a kinetic energy of 80 eV impinging on silicon. A) trajectories of the impinging silicon ions, b) depth distribution of the impinging silicon ions.**

By measuring the effective dopant concentration during ARP e.g. using capacitance voltage (CV) measurements a 2 to 1 correlation with the E1 concentration was found. This means that for each appearing E1 defect two free holes in the p-type silicon are removed. The interpretation of that measurement is that for each negatively charged boron acceptor atom, which undergoes a defect reaction, a positively charge donor defect is generated.

The current knowledge[24] of the formation of the $B_iO_i$ defect during irradiation is as follows: High energy particles of the irradiation generate interstitial silicon atoms and vacancies in the silicon lattice. EPR investigations done by George Daniels Watkins were not able to reveal a signal related to interstitial silicon after irradiation of silicon at low temperatures of about 20K.[25] Rather EPR signals related to e.g. aluminum or boron appear at that low temperature. These findings proof that the silicon interstitial is a very fast diffuser even at temperatures of about 20K under electron irradiation conditions. The silicon interstitial moves due to its high mobility amongst others to substitutional acceptors and kicks them out of their lattice position at 20K due to the Watkins replacement reaction. After the Watkins replacement reaction these substitutional acceptors are defects which reside on interstitial lattice positions and are denoted e.g. by $B_i$. By further increasing the temperature to about 250K the EPR signal of the secondary defect interstitial boron $B_i$ disappears as well. This is the proof that this defect is also a fast diffuser. Finally, the highly mobile interstitial boron $B_i$ finds an interstitial oxygen atom $O_i$ and forms the $B_iO_i$ defect, which is stable at room temperature.

This model has been accepted over the past years by numerous authors. Interestingly, to the best of our knowledge no one has published a reproduction of these important low temperature irradiation experiments.[26]

Remaining questions are: What happens at these low temperatures during irradiation in the silicon? Are there different approaches to explain the experimental observations, which do not need the assumption of fast diffusion and a $Si_i$ $B_{Si}$ position exchange (Watkins replacement mechanism) at 20K?

The experiments done at that time[25] use 1-3 MeV electrons, which impinge on silicon at temperatures around 20K. The detailed experimental setup is to our knowledge not described in the literature. A current density of about 2.5 µA/cm² is sometimes being reported[27] but the duration of that irradiation is rarely discussed. If we assume an electron energy of 3 MeV the primary knock-on silicon atom (PKA) will, calculated with the relativistic formula reported by Akkerman et al.,[28] end up with an energy of about 930eV. Using the versatile software tool SRIM[29] the movement of such generated interstitial silicon atoms can be visualized by simulation. In that special case the maximum distance of movement of such PKAs is about 10 nm. A lower limit can be estimated as well: For 1 MeV electron irradiation about 10% of the silicon recoil atoms end up with kinetic energies larger than 80 eV if we follow the calculations of Leroy and Rancoita.[30] For interstitial silicon atoms with this energy SRIM calculates the peak maximum of the depth distribution to about 10 Å (see Figure 3). Hence, we can assume that about 50% of these interstitial silicon atoms travel distances of more than 1 nm after the collision event with an electron. The minimal distance that a PKA must travel to reach a boron atom in boron doped silicon with a boron density of $10^{16}$cm$^{-3}$ is 23 nm assuming a homogenous boron distribution. If we assume a microscopic homogenous distribution of collision events then every 23rd of these more than 1 nm travelling interstitial silicon atoms have the possibility to reach a boron atom. Assuming one hour of irradiation with above current density we end up with about $6\times10^{17}$ electrons per cm$^2$. If these electrons are all absorbed in silicon of 1 mm thickness[31] we have $6\times10^{18}$ electrons per cm$^3$ generating PKAs. If we subsequently consider all our assumptions (10% of PKA have more than 80 eV kinetic energy, 50% of these travel more than 1 nm and every 23th of these is then able to reach a boron atom) we end up with $1.3\times10^{16}$ interstitial silicon atoms per cm$^3$ which are able to reach a boron atom due to their energy gained from the 1 MeV electron irradiation. If additionally assumed that only 10% of these interstitial silicon atoms are finally captured by the substitutional boron this would give a density of about $10^{15}$ so called "interstitial boron" $B_i$ per cm$^3$, which is an EPR detectable value at that time.[26]

From this point of view, it is not necessary to assume a thermal or an athermal fast diffusion for the interstitial silicon, which may be even enhanced by changing the charge state of the defect during electron irradiation due to charge carrier capture and emission processes.[32] It could be possible that some of the boron atoms capture interstitial silicon atoms, which move through the lattice due to the collision event. The attraction of both might be enhanced due to the lattice deformation caused by the substitutional boron atom. A coulombic attraction at these low temperatures seems on the first view unlikely although recent DFT calculations show energetical favorability of the B-Si-pair formation.[33] A $B_{Si}$-$Si_i$-defect would then have been formed, which has in its ground state $C_{3v}$ symmetry. Due to illumination this configuration may change to $C_{1h}$ symmetry, for which DFT calculations show indications that this configuration is EPR active.[33]

The explanation of the ARP by an $A_{Si}$-$Si_i$-defect mode is now as follows. Similar processes as described above would happen if irradiation takes place at room temperature. The effectiveness of $B_{Si}$-$Si_i$-defect formation would be even enhanced due to the coulombic attraction of the now charged defects. In the ground mode the $B_{Si}$-$Si_i$-defect with $C_{3v}$ symmetry is a donor.[10,33] Hence, the E1 related activation energy of a charge state change observed in DLTS measurements and the ARP can be explained by the formation of this $B_{Si}$-$Si_i$-defect mode during irradiation.

An interesting route to follow for further investigations would be to assume that the AR and LID phenomenon in boron doped silicon are caused by the same defect complex namely the $B_{Si}$-$Si_i$-defect. One interesting example for this assumption is the observation of the instability of the ARP under charge carrier injection by Nitescu et al.[34] They measured activation energies for the forward and backward reaction of the defect related to the ARP, which are not so far away from activation energies measured for the forward and backward reaction of the so-called BO-LID as investigated by Bothe et al.[18]

Following this understanding, it might be possible to remove the interstitial silicon atom from the substitutional boron atom and store it somewhere else by a sophisticated treatment. The consequence of that would be a restoring of the acceptor properties of boron and hence a solution to the ARP problem of the LGADs.

## 4. Summary and conclusion

Low gain avalanche detectors (LGAD), which are fabricated at CiS in Erfurt, Germany, were shown to be prone to the acceptor removal phenomenon (ARP) under irradiation, which leads to the loss of their ability to intrinsically amplify a signal. A careful reexamination of the basic physics during MeV electron irradiation of silicon allows the conclusion that the assumption of a fast diffusion of interstitial silicon at low temperatures is not necessary. The energy of the generated interstitial silicon gained due to the collision with the incident electron is sufficient enough to explain the movement of interstitial silicon to sinks like the substitutional boron without diffusion. In that way the generation of the $B_{Si}$-$Si_i$-defect during irradiation can be understood. If the $B_{Si}$-$Si_i$-defect has been formed during irradiation, then it is in its ground mode a donor and hence, can explain the ARP. First LTPL results of a PL feature called S line in boron doped silicon are reported to be related to the LID phenomenon, which might be also caused by the $B_{Si}$-$Si_i$-defect. Based on these findings it seems to be worth to reconsider the knowledge on defect formation and defect identification in silicon under irradiation.

## 5. Acknowledgement

This work was supported by the project SimASiSii of the Deutsche Forschungsgemeinschaft (DFG project number 445152322, KR 2228/11-1, LA 4623/1-1, RU 1383/6-1) and the project eLGAD of the BMWi (49MF220095).